\documentclass[aip,jap,reprint]{revtex4-1}

\usepackage{color}
\usepackage{soul}
\usepackage{setspace}
\usepackage{gensymb}
\usepackage{graphicx}
\usepackage{epstopdf}
\usepackage{amssymb}
\usepackage[colorlinks=true,linkcolor=blue]{hyperref}
\usepackage{array}
\usepackage{bm}
\usepackage{float}
\usepackage{amsmath}

\newcommand{\JCo}{\ensuremath{ J_{1}^{\rm{Co}} }}
\newcommand{\JJCo}{\ensuremath{ J_{2}^{\rm{Co}} }}

\newcommand{\tCo}{\ensuremath{ \theta^{\rm{Co}} }}
\newcommand{\tFe}{\ensuremath{ \theta^{\rm{Fe}} }}

\newcommand{\JFe}{\ensuremath{ J_{1}^{\rm{Fe}} }}
\newcommand{\JJFe}{\ensuremath{ J_{2}^{\rm{Fe}} }}

\usepackage[markup=underlined]{changes}
\definechangesauthor[color=red]{Erol}
\definechangesauthor[color=green]{Zach}
\definechangesauthor[color=blue]{Paul}

\usepackage{todonotes}
\makeatletter
\setremarkmarkup{\todo[color=Changes@Color#1!20,size=\scriptsize]{#1: #2}}
\makeatother

\begin{document}
	
	%\title{Non-collinear coupling across RuCo and RuFe alloys}
	\title{Non-collinear coupling across RuCo and RuFe alloys}

	\author{Z. R. Nunn}%
	\email{znunn@sfu.ca}%Lines break automatically or can be forced with \\
	\author{E. Girt}%
	\email{egirt@sfu.ca}
	\affiliation{Simon Fraser University, 8888 University Drive, Burnaby, British Columbia V5A 1S6, Canada}
	
	\date{\today}% It is always \today, today,
	
	\begin{abstract}

	Spintronic applications, which rely on spin torques for operation, would greatly benefit from a non-collinear alignment between magnetizations of adjacent ferromagnetic layers for maximum performance and reliability. We demonstrate that such an alignment can be created and controlled by coupling two ferromagnetic layers across magnetic coupling layers. These coupling layers consist of a non-magnetic material, Ru, alloyed with ferromagnetic elements of Co or Fe. Changing the composition and thickness of the coupling layer enables control of the relative angle between the magnetizations of ferromagnetic layers between 0$\degree$ and 180$\degree$. The onset of the non-collinear alignment between ferromagnetic layers coincide with the advent of magnetic order in the coupling layer. This study will map the range of concentrations and thicknesses of RuCo and RuFe coupling layers that give rise to non-collinearity between Co layers.

	\end{abstract}
	
%	\pacs{Valid PACS appear here}% PACS, the Physics and Astronomy
%	% Classification Scheme.
%	\keywords{Antiferromagnetic Coupling; Synthetic Antiferromagnet; Bilinear; Biquadratic;}%Use showkeys class option if keyword
%	%display desired
	\maketitle

%Zach
	 Spintronic based applications emerged and have been under heavy investigation since the early 2000's as giant/tunneling magnetoresistance (GMR/TMR) effects$^{1,2,3}$ combined with spin transfer torque (STT)$^{4}$ enabled both the detection and manipulation of magnetic moment orientation with an electrical current. Two spintronic devices important for the advancement of electronics industry are spin transfer torque magnetic random access memory (STT-MRAM)$^{5}$ and spin torque nano-oscillator (STNO) devices$^{6,7}$. In these devices the magnitude of STT is represented by the double cross product $M_{1}\times (M_{1}\times M_{2})$$^{4}$, where $M_{1}$ and $M_{2}$ are the unit vectors of the magnetic moments of adjacent magnetic layers. When the magnetic moments $M_{1}$ and $M_{2}$ have collinear alignment the STT is zero. This has created a problem for the performance and reliability of STT devices as thermal fluctuations or external magnetic fields are relied upon to provide a non-collinear alignment between $M_{1}$ and $M_{2}$. For this reason a non-collinear alignment of the magnetic moments in neighboring ferromagnetic layers is desired for the optimal performance of these devices$^{8,9,10}$. 

	The discovery of interlayer exchange coupling in 1986$^{11}$ allowed for the first time, control of antiferromagnetic coupling between two ferromagnetic films. The coupling was observed across the majority of 3d, 4d, and 5d non-magnetic metallic spacer layers$^{11,12,13}$. The interlayer exchange coupling was found to oscillate between antiferromagnetic ($\theta$ = 180$\degree$) and ferromagnetic ($\theta$ = 0$\degree$) states as a function of the non-magnetic spacer layer thickness, where $\theta$ is the angle between the magnetic moments of the ferromagnetic layers.  Unfortunately, the transition from antiferromagnetic to ferromagnetic alignment occurs within a very narrow spacer layer thickness range which prevented the control of $\theta$ between 0$\degree$ and 180$\degree$.	
	
	This article presents novel coupling materials which allow for non-collinear coupling between two ferromagnetic layers. Furthermore, it was found that by varying the coupling layer thickness and/or composition, precise control of the non-collinear coupling angle and the coupling strength can be achieved. The coupling layer consists of a non-magnetic material, which on its own can provide antiferromagnetic coupling between ferromagnetic layers, alloyed with ferromagnetic elements with a composition close to the non-magnetic/magnetic transition concentration of the alloy. 
	
	The non-magnetic material, Ru, is sandwiched between Co magnetic layers and alloyed with magnetic elements of Co and Fe to create Co$|$RuCo$|$Co and Co$|$RuFe$|$Co structures. RuCo and RuFe coupling layers are good candidates for this study as Co and Fe form solid solutions with Ru over a large composition range$^{14,15}$. It is the competition between the antiferromagnetic coupling through Ru and the ferromagnetic coupling through Co or Fe that results in non-collinear coupling. To date, paramagnetic and antiferromagnetic materials have been used to achieve coupling between ferromagnetic layers. The coupling materials presented in this work can have a magnetic moment larger than that of ferromagnetic Ni. It will be shown that non-collinear coupling exists in a wide composition range of RuCo and RuFe, and that the angle between Co layers can be precisely controlled by the composition and thickness of the coupling layer.

	\section{Non-collinear Coupling thickness and composition dependence}

 	\begin{figure*}[ht]
 		\centering
 		\includegraphics[width=0.99\textwidth]{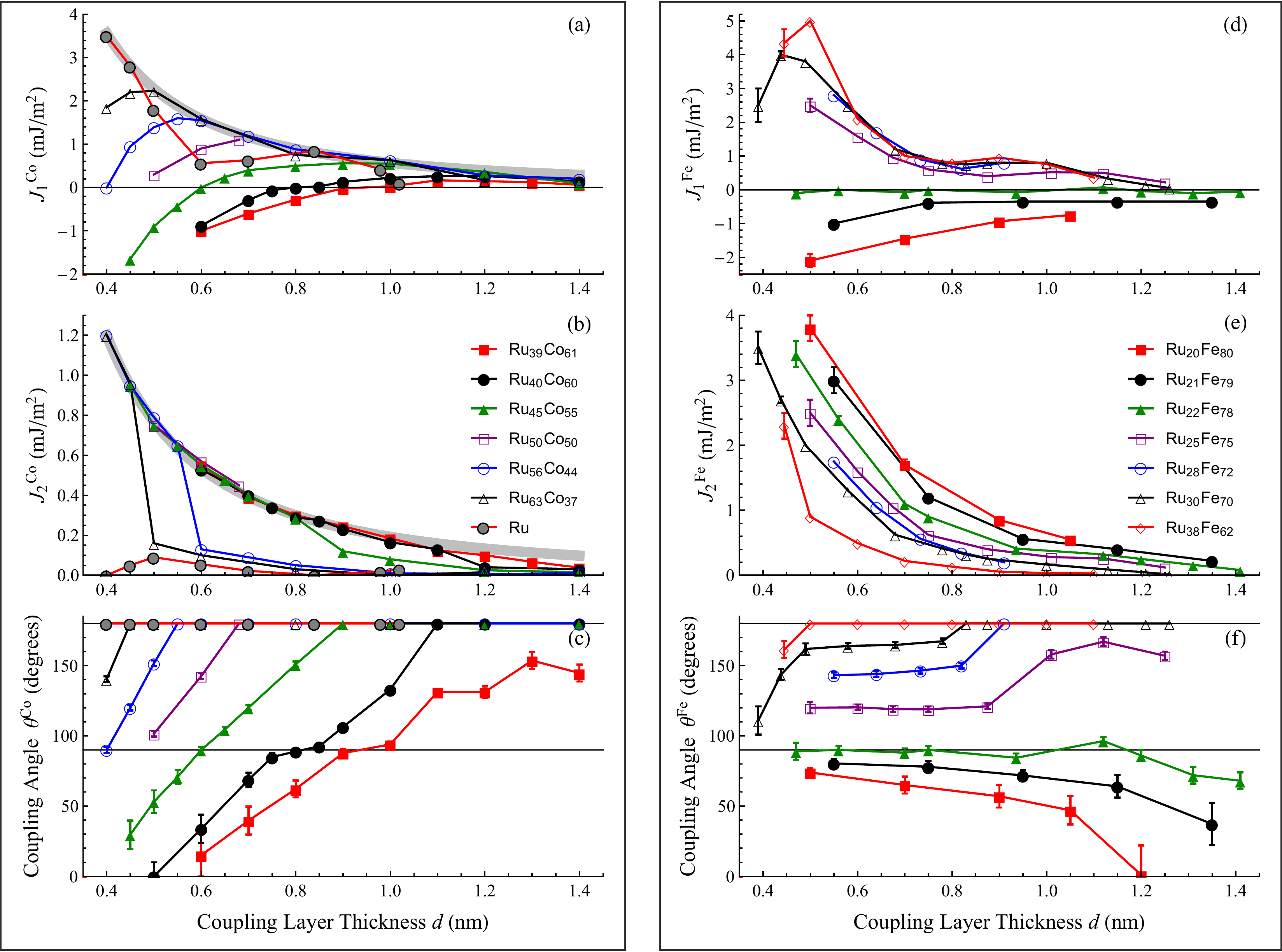}
 		\caption{ (left) (a) $\JCo$, (b) $\JJCo$, and (c)  $\tCo$ of Co(2 nm)$|$Ru$_{100-x}$Co$_{x}$($d$)$|$Co(2 nm)  for $0.4 \leq d \leq 1.4$ nm and $37 \leq x \leq 61$ at.$\%$ at 298 K. (right) (d) $\JFe$, (e) $\JJFe$, and f) $\tFe$ of Co(2 nm)$|$Ru$_{100-x}$Fe$_{x}$($d$)$|$Co(2 nm) for $0.4 \leq d \leq 1.4$ nm and $62 \leq x \leq 80$ at.$\%$ at 298 K. A broad grey $1/d^{2}$ dependence, and $J_{1}$ and $J_{2}$ coupling data of Co(2 nm)$|$Ru($d$)$|$Co(2 nm) are added for comparison in (a) and (b). $J_{1}$, $J_{2}$, and $\theta$ data for coupling of Co layers across Ru$_{100-x}$Co$_{x}$ $x < 37$ and Ru$_{100-x}$Fe$_{x}$ $x < 62$ have been omitted as non-collinear coupling was not observed for $0.4 \leq d \leq 1.4$ nm.}
 		\label{fig:J1J2phi}
 	\end{figure*}
	
	This section presents one of the main results of this article, controllability of the non-collinear coupling strength and angle. Two structures were investigated. First, Co(2 nm)$|$Ru$_{100-x}$Co$_{x}$($d$)$|$Co(2 nm) with the bilinear $(\JCo)$ and biquadratic ($\JJCo$) coupling strengths and coupling angle ($\tCo$) presented in Fig.\ref{fig:J1J2phi} (a), (b) and (c), respectively. The thickness of the coupling layer is $0.4 \leq d \leq 1.4$ nm and the Co concentration of $37 \leq x \leq 61$ at.$\%$. Second, Co(2 nm)$|$Ru$_{100-x}$Fe$_{x}$($d$)$|$Co(2 nm) with $\JFe$,  $\JJFe$ and $\tFe$ displayed in Fig.\ref{fig:J1J2phi} (d), (e) and (f) with coupling layer thickness $0.4 \leq d \leq 1.4$ nm and Fe concentration of $62 \leq x \leq 80$ at.$\%$.
	
	For the readers convenience the phenomenological description of non-collinear coupling is reviewed; the angle between magnetic moments of two ferromagnetic layers is between 0$\degree$ and 180$\degree$. The coupling energy, which determines the relative orientation between ferromagnetic moments, can be described as the superposition of $J_1$ and $J_2$ coupling terms in $E_{\text{coupling}}=J_{1}cos(\theta)+J_{2}cos^{2}(\theta)$, where $\theta$ is the angle between \ magnetic moments of ferromagnetic layers$^{16}$. The bilinear term favours a collinear alignment. In this model $J_1 < 0$ is defined as ferromagnetic (0$\degree$) and $J_1 > 0$ is antiferromagnetic (180$\degree$) alignment. The biquadratic term favours an orthogonal alignment (90$\degree$) when $J_2$ is positive, as observed in all presented measurements. Non-collinear coupling occurs when $J_2 > |J_1|/2$. The values of $J_1$, $J_2$ and $\theta$ are determined by fitting the magnetization, $M$, as a function of field, $H$, of the coupled structure$^{17}$, as described in the methods section. 

	As shown in Fig.\ref{fig:J1J2phi} the bilinear coupling strength and coupling angle of Co$|$Ru$(d)|$Co, for $0.4 \leq d \leq 1.02$, follows the expected thickness dependence$^{18}$, while the biquadratic coupling strength is small. With the introduction of Co into the Ru coupling layer, a loss of the oscillatory dependence of $\JCo$ with $d$ is observed, and little to no change of $\JJCo$. However, at some critical concentration of Co, $\JJCo$ shows a rapid increase which coincides with the onset on non-collinear coupling, see Fig. \ref{fig:J1J2phi} (b) and (c). This occurs when the condition for non-collinear coupling is met, $J_2 > |J_1|/2$. For large concentrations of Co, $x \geq 55 $, $\JCo$  crosses into the ferromagnetic region  while $\JJCo$ approaches a $1/d^2$ dependence, see grey line in Fig.\ref{fig:J1J2phi} (b). The smooth non-oscillatory thickness dependence of $\JCo$ shown in Fig.\ref{fig:J1J2phi} (a) could be due to the superexchange background predicted for an insulating$^{19,20}$ and metallic$^{21,22,23}$ coupling layers.

	In order to understand the previous results, identical measurements were performed for a RuFe coupling layer were $\JFe$, $\JJFe$ and $\tFe$ are shown in Fig \ref{fig:J1J2phi} (d), (e), and (f), respectively. In contrast to $\JCo$, $\JFe$ does not show a reduction of oscillations when Fe is introduced into Ru. This suggests that the reduction of oscillations of $\JCo$ could be due to a decrease of the Co gradient at the Co$|$RuCo interface. Adding Fe to Ru does not effect the Co gradient at the Co$|$RuFe interface, and similarly to $\JJCo$, $\JJFe$ monotonically decreases with $d$. Additionally, it is clear that the coupling strength of both the bilinear and biquadradic terms are material dependent since $\JCo<\JFe$ and $\JJCo<\JJFe$. The largest measured value for orthogonal coupling is $\JJFe$ = 3.4 mJ/m$^2$ for Co$|$Ru$_{22}$Fe$_{78}$(0.48 nm)$|$Co. This is over twice as large as any $J_2$ value previously reported$^{24}$.
	
	The angle $\tCo$, between magnetic moments of the Co layers in Co$|$Ru$_{100-x}$Co$_x$($d$)$|$Co structures changes with both $d$ and $x$. Fig.\ref{fig:J1J2phi} (c) shows that for all RuCo compositions, $\tCo$ increases with $d$. The concentration of Co across the coupling layer is expected to decrease from the Co$|$RuCo interfaces to the center of the coupling layer due to inter-diffusion, as experimentally verified for Co$|$Ru$|$Co structures$^{25}$. As a result, the average concentration of Co in Ru$_{100-x}$Co$_x$ will scale with $d$, meaning a smaller $d$ will have a larger $x$. This in part is responsible for an increase of $\tCo$ at higher $d$. Additionally, the reduced Co material gradient may account for the reduction of the $\theta$($d$) slope in Fig.\ref{fig:J1J2phi} (c).
	
	The reader's attention should now be turned to Fig.\ref{fig:J1J2phi} (f), in the thickness region of $0.5 \leq d \leq 0.8$ nm. Within this region the coupling angle remains constant with $d$, for most of the RuFe alloys. However, varying the concentration allows for a control of the coupling angle between 180$\degree$ and 90$\degree$. From a fabrication point of view, this is very important in order to utilize non-collinear coupling in applications since small thickness errors do not contribute to changes in coupling angle. 
	
	\section{Magnetic Properties of Non-collinear Coupling Layers} 	

		\begin{figure*}[ht!]
			\centering
			\includegraphics[width=0.99\textwidth]{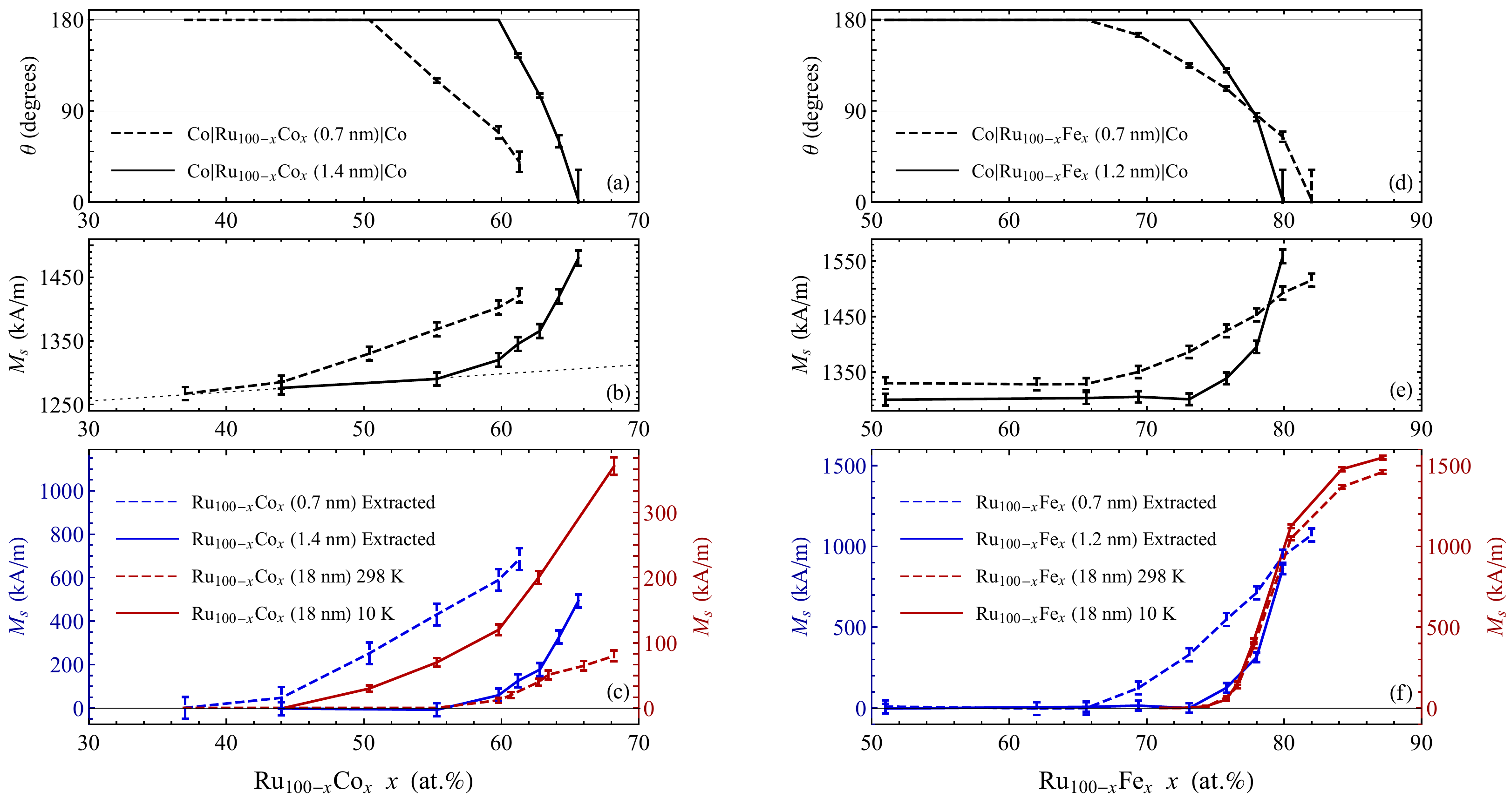}
			\caption{(a) $\tCo$, and (b) $M_s$ of Co(2 nm)$|$Ru$_{100-x}$Co$_{x}$(0.7 nm)$|$Co(2 nm) and Co(2 nm)$|$Ru$_{100-x}$Co$_{x}$(1.4 nm)$|$Co(2 nm) at 298 K. (c) The extracted $M_s$ of Ru$_{100-x}$Co$_{x}$(0.7 nm) and Ru$_{100-x}$Co$_{x}$(1.4 nm) at 298 K and measured $M_s$ of Ru$_{100-x}$Co$_{x}$(18 nm) at 10 K and 298 K. (d) $\tFe$, and (e) $M_s$ of Co(2 nm)$|$Ru$_{100-x}$Fe$_{x}$(0.7 nm)$|$Co(2 nm) and Co(2 nm)$|$Ru$_{100-x}$Fe$_{x}$(1.2 nm)$|$Co(2 nm) at 298 K. (f) The extracted $M_s$ of Ru$_{100-x}$Fe$_{x}$(0.7 nm) and Ru$_{100-x}$Fe$_{x}$(1.2 nm) at 298 K and measured $M_s$ of Ru$_{100-x}$Fe$_{x}$(18 nm) at 10 K and 298 K. In (c) two $M_s$ scales correspond to Ru$_{100-x}$Co$_{x}$(0.7 nm) and Ru$_{100-x}$Co$_{x}$(1.4 nm) extracted (left), and Ru$_{100-x}$Co$_{x}$(18 nm) (right). Data extending to $x=0$ not shown in (b) and (c) are in Extended Fig.\ref{fig:MsAll}.}
			\label{fig:MsNCL}
		\end{figure*}
		
		This section presents data on the magnetic properties of the coupling layer and relates them to the region of data for which non-collinear coupling occurs. The data behind the main ideas of this section are presented in Fig.\ref{fig:MsNCL}. In order to understand how non-collinear coupling relates to the magnetic moment of the coupling layer, three different samples were studied: Co$|$Ru$_{100-x}$Co$_x$($0.7$ nm)$|$Co, Co$|$Ru$_{100-x}$Co$_x$($1.4$ nm)$|$Co and a bulk Ru$_{100-x}$Co$_x$($18$ nm) without any adjacent Co layers on either side. Fig.\ref{fig:MsNCL} (a) and (b) show the coupling angle and the saturation magnetization, $M_{s}$, of  Co$|$Ru$_{100-x}$Co$_x$($0.7$ nm)$|$Co, and  Co$|$Ru$_{100-x}$Co$_x$($1.4$ nm)$|$Co. Fig.\ref{fig:MsNCL} (c) compares the magnetic properties of the Ru$_{100-x}$Co$_x$($18$ nm), to the extracted magnetic properties of Ru$_{100-x}$Co$_x$($0.7$ nm) and Ru$_{100-x}$Co$_x$($1.4$ nm) from Fig.\ref{fig:MsNCL} (b); see further discussion below. In order to contrast how the magnetic species influence the magnetic properties of the coupling layer the above was repeated substituting RuCo with RuFe in Co$|$Ru$_{100-x}$Fe$_x$($0.7$ nm)$|$Co and Co$|$Ru$_{100-x}$Fe$_x$($1.2$ nm)$|$Co and Ru$_{100-x}$Fe$_x$($18$ nm); see Fig.\ref{fig:MsNCL} (d), (e), and (f). 
		
		A monotonic increase in $M_{s}$ is observed for all Co$|$Ru$_{100-x}$Co$_x|$Co and Co$|$Ru$_{100-x}$Fe$_x|$Co structures with increasing concentration of Co or Fe, respectively; see Fig.\ref{fig:MsNCL} (b) and (e). For the Co$|$Ru$_{100-x}$Co$_x|$Co structure, the $M_{s}$ increases linearly with $x$ for $0 \leq x \leq 44$. This linear increase of the $M_s$ with $x$ is represented by a dotted line in Fig.\ref{fig:MsNCL} (b) and can be seen more clearly in Extended Fig.\ref{fig:MsAll}. This can be explained by the increase of $M_s$ of the interface atoms of the Co layers$^{17}$. It has been shown that in Co$|$Ru multilayers, Co atoms, at the Ru interface have a reduced magnetic moment. Adding Co to Ru is thus, expected to modify the electronic environment of the ferromagnetic Co layer's interface atoms, and increase their magnetization. For $x>44$	the $M_s$ of Co$|$Ru$_{100-x}$Co$_x$($d$)$|$Co starts to sharply increase with $x$ and deviate from the dotted line. This sharp increase is attributed to the onset of a magnetic order of the RuCo coupling layer. Evidence of this can be observed from the data in Fig.\ref{fig:MsNCL} (c), which compares the magnetic moment of Ru$_{100-x}$Co$_x$($18$  nm) to  Ru$_{100-x}$Co$_x$($0.7$ nm) and Ru$_{100-x}$Co$_x$($1.4$ nm). It can be observed that the onset of magnetic order for the Ru$_{100-x}$Co$_x$($1.4$ nm) and Ru$_{100-x}$Co$_x$($18$ nm) occurs at the same Co concentration, $x=60$. The magnetic moment of Ru$_{100-x}$Co$_x$($0.7$ nm) and Ru$_{100-x}$Co$_x$($1.4$ nm) is extracted from Fig.\ref{fig:MsNCL} (b) by subtracting the sharp increase in $M_s$ from the linear background and normalizing by the thickness of the coupling layer with respect to the total thickness of  Co$|$Ru$_{100-x}$Co$_x$($d$)$|$Co. It is important to identify that for Co$|$Ru$_{100-x}$Co$_x$($1.4$ nm)$|$Co the  non-collinear coupling and magnetic order occur in the same concentration range, $60<x<66$. However, for a thinner coupling layer, Co$|$Ru$_{100-x}$Co$_x$($0.7$ nm)$|$Co, both the non-collinear coupling region and the region of magnetic order broadens and shifts to lower concentrations. This is attributed to inter-diffusion of Co, resulting in thinner Ru$_{100-x}$Co$_x$ having higher $x$. In addition, the magnetic proximity effect could be polarizing the entire coupling layer below a certain thickness $d$, which could also be responsible for the observed broadening and shift of the non-collinear coupling region. From the $M_s$ dependance on $x$ of Ru$_{100-x}$Co$_{x}$(18 nm) performed at 10 and 298 K in Fig.\ref{fig:MsNCL} (c), it is evident that Ru$_{100-x}$Co$_{x}$ exhibit superparamagnetic behaviour for $44<x<60$. This could explain the different magnitude of $M_s$ values when comparing the extracted values to that measured from the bulk as proximity effects may be polarizing the coupling layers with a greater intensity then at which the bulk was measured ($>7$ T). 
		
		The measurements of the $\theta$ and $M_s$ of Co$|$Ru$_{100-x}$Fe$_x$($0.7$ nm)$|$Co and Co$|$Ru$_{100-x}$Fe$_x$($1.2$ nm)$|$Co as a function of $x$ are shown in Fig.\ref{fig:MsNCL} (d), and (e). The Extended Fig.\ref{fig:MsAll} shows that the $M_s$ of Co$|$Ru$_{100-x}$Fe$_x$($d$)$|$Co has an increase of $M_s$ for $0 < x \leq 36$ and is constant for $36<x<65$. Similarly to the linear increase in $M_s$ of the RuCo coupling layer, the increase of $M_s$ is not associated with the RuFe layer, but is related to the CoFe atoms at the Co$|$RuFe interfaces$^{17}$. With further increase in $x$ a second sharp increase in $M_s$ occurs which coincides with the onset of the non-collinear coupling, Fig.\ref{fig:MsNCL} (d). Similarly to RuCo, this $M_s$ increase relates to the onset of a magnetic order of the RuFe coupling layer, as observed by comparing the $M_s$ of Ru$_{100-x}$Fe$_x$($1.2$ nm) to Ru$_{100-x}$Fe$_x$($18$ nm); see Fig.\ref{fig:MsNCL} (f). Furthermore, the magnitude of $M_s$ for Ru$_{100-x}$Fe$_x$($1.2$ nm) and Ru$_{100-x}$Fe$_x$($18$ nm) coincide over the full concentration range. Once again, it is observed that non-collinear coupling and magnetic order for the Co$|$Ru$_{100-x}$Fe$_x$($1.2$ nm)$|$Co occur in the same concentration range, $73<x<80$. Lastly, for Co$|$Ru$_{100-x}$Fe$_x$($0.7$ nm)$|$Co a shift and broadening occurs for the concentration range over which non-collinear coupling and the magnetic ordering occurs. This can be attributed to the proximity polarization effect and possible inter-diffusion at Co$|$RuFe interfaces. 
		
		It is important to point out that these coupling layers can have large $M_s$ values. The orthogonal alignment ($\theta=90\degree$) in Co$|$Ru$_{42}$Co$_{58}$(0.7 nm)$|$Co and Co$|$Ru$_{22}$Fe$_{78}$(0.7 nm)$|$Co was achieved across a coupling layer with $M_s$ = 500, and 700 kA/m, respectively. These coupling layers have larger $M_s$ then that of ferromagnetic Ni, $M_{s}$ = 488 kA/m$^{26}$. This is the first demonstration of coupling occurring across a magnetic material; only paramagnetic and antiferromagnetic layers have been used to provide interlayer exchange coupling.

	\section{Non-collinear Coupling Mechanism}
	
	Biquadratic coupling, $J_2$, controls the non-collinear alignment of magnetic layers. Thus, this section will give a brief review of the known biquadratic coupling mechanisms and focus on the spatial fluctuation mechanism as it could be responsible for the $J_2$ in this article. $J_2$ can originate from intrinsic and extrinsic sources$^{27,28}$. In the studied structures, the measured $J_2$ always favours a perpendicular alignment and has a strength comparable to the bilinear coupling, suggesting an extrinsic source$^{16}$. Extrinsic sources of biquadratic coupling could be due to uncorrelated film roughness$^{16}$, pin-holes$^{29}$, loose spins$^{30}$ and spatial fluctuations$^{31}$. First, film roughness, pin-holes, and the loose spin mechanism will be briefly discussed followed by a more in-depth discussion of the spatial fluctuation mechanism as it best relates to the samples in this article.
	
	In films with uncorrelated film roughness, the biquadratic coupling contribution can be estimated from a simple model that assumes one smooth ferromagnetic/coupling layer interface and the other interface with a corrugation described by a sinusoidally varying function$^{16}$. From atomic force microscopy, the root mean square roughness of the samples in this article was determined to be $\sim 0.1$ nm. From the exchange stiffness of Co ferromagnetic layers$^{17}$, 1.2$\times 10^{-11}$ J/m, the estimated upper limit of $J_2$ = 0.15 mJ/m$^2$ could exist. This is an order of magnitude smaller than the largest measured $\JJCo$ and $\JJFe$; see Fig.\ref{fig:J1J2phi}. The loose spin model assumes that the coupling layer contains magnetic atoms that are in a paramagnetic state. As shown in Fig.\ref{fig:MsNCL} the RuCo and RuFe coupling layers have magnetic properties and therefore cannot be described by this model. Fe and Co are soluble over a large composition range in Ru, thus, the existence of pin-holes in our films is not expected. Microstructures of Co$|$Ru multilayers have been extensively studied in both academia$^{32}$ and industry$^{33}$ and the presence of pin-holes have not been reported.

	The spatial fluctuation mechanism is based on an magnitude change of the bilinear coupling term, $J_1$, across the films plane. These fluctuations are usually due to a spatial variation of the coupling layer thickness$^{34,35}$. The competition between the coupling and stiffness energies leads to an orthogonal (90$\degree$) alignment between the magnetic moments. Following the theory proposed by Slonczewski$^{31}$ the strength of the biquadratic coupling between two identical magnetic layers of thickness, $t$, across a coupling layer can be determined from the following relation,
	\begin{equation}
	J_{2} = \frac{4 (\Delta J_1)^2 L}{\pi^3 A_{ex}} \coth(\pi \frac{t}{L})
	\label{eq:Eq.1}
	\end{equation}
	where $A_{ex}$ is the exchange stiffness of the Co layers, and $\Delta J_1$ and $L$ are the magnitude change and length of spatial fluctuations of $J_1$ across the films surface, respectively.

		\begin{figure}[tbph]
			\centering
			\includegraphics[width=0.49\textwidth]{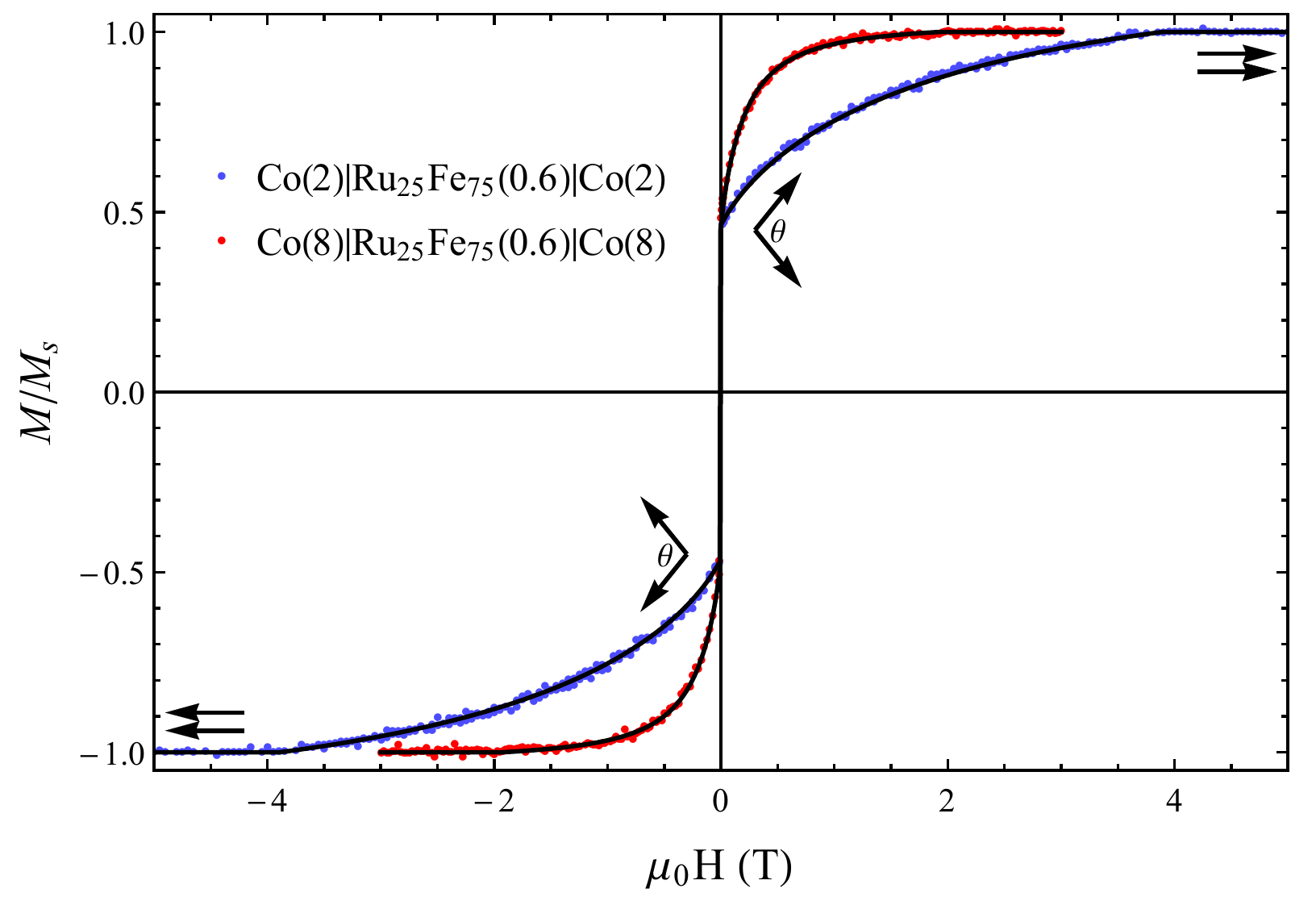}
			\caption{
				Solid circles represent $M(H)$ measurements of Co($t$)$|$Ru$_{25}$Fe$_{75}$(0.6 nm)$|$Co($t$), where $t$ is 2 and 8 nm. Solid lines are simulations of the $M(H)$ data used to determine $\JFe$ and $\JJFe$ as described in the methods section. Arrows illustrate the directions of the magnetic moments of the Co layers in zero and large external magnetic fields.}
			\label{fig:MHCurves2}
		\end{figure}		 
	 
	The structures in this article can have spatial fluctuations, $\Delta J_1$, originating from the randomness of the distribution of magnetic atoms in RuCo and RuFe coupling layers. As an example, this can be visualized in Co$|$Ru$_{25}$Fe$_{75}|$Co where ferromagnetic Co layers are separated by a two monolayer thick magnetic Ru$_{25}$Fe$_{75}$ coupling layer. In this structure atoms in the Co layer are randomly ferromagnetically coupled across a pair of Fe atoms, antiferromagnetically coupled across a pair of Ru atoms, or antiferromagnetically coupled across one Ru and one Fe atom.  Since the atoms in Ru$_{25}$Fe$_{75}$ are homogeneously distributed, it is expected that the size of spatial fluctuations, $L$, is very small, approaching atomic distances. From Eq.\ref{eq:Eq.1}, if $L$ is less than the magnetic layer thickness $t$, $\coth(\pi \frac{t}{L})=1$, and thus, $J_{2}$ is independent on $t$.  Fig.\ref{fig:MHCurves2} shows $M(H)$ measurements of  Co($t$)$|$Ru$_{25}$Fe$_{75}$(0.6 nm)$|$Co($t$) for two Co layer thicknesses $t$ = 2 and 8 nm. It was found that both structures have  the same coupling strength of $\JFe$ = 1.65 mJ/m$^2$, $\JJFe$ = 1.45 mJ/m$^2$ and a zero magnetic field coupling angle $\tFe$ = $118\pm2^{\degree}$. Due to $J_{2}$ being independent of $t$, the spatial fluctuations, $L$, in Eq.\ref{eq:Eq.1} must be less than 2 nm. This is promising for nanometer size devices as 2 nm is well below the current lithography process node size used in the fabrication of spintronic devices.	

	\section{Conclusion}	
	
	This article presented a systematic study of non-collinear coupling of ferromagnetic Co layers across RuCo and RuFe coupling layers. It was shown that the angle between magnetic moments of ferromagnetic Co layers can be controlled by varying the thickness and/or composition of the RuCo and RuFe layers. For optimal non-collinear angle control, the concentration, $x$, and thickness, $d$, ranges are $44 \leq x \leq 61$ at.$\%$ and $0.4 \leq d \leq 1$ nm for Co$|$Ru$_{100-x}$Co$_{x}$($d$)$|$Co structures, and $68 \leq x \leq 82$ at.$\%$ and $0.5 \leq d \leq 0.8$ nm for Co$|$Ru$_{100-x}$Fe$_{x}$($d$)$|$Co structures. It was shown that non-collinear coupling is correlated with a magnetic order of the coupling layer with saturation magnetizations up to 700 kA/m for RuCo and 1000 kA/m for RuFe. The non-collinear coupling mechanism is interpreted to originate from bilinear coupling spatial fluctuations. In addition to controlling relative magnetizations with a coupling layer, the strength of coupling was found to be extremely large which will allow for new non-collinear spintronic designs with magnetic moments tilted in respect to the film plane.

	\bibliography{mybib}

	\section{References}

\small

$^{1}$ Binasch, G., Grunberg, P., Saurenbach, F., Zinn, W. Enhanced magnetoresistance in layered magnetic structures with antiferromagnetic interlayer exchange. Phys. Rev. B \textbf{39}, 4828 (1989).

$^{2}$ Baibich, M. N., Broto, J. M., Fert, A., Nguyen van Dau, F., Petro, F., Eitenne, P., Creuzet, G., Friederich, A., Chazelas, J., Giant Magnetoresistance of (001)Fe|(001)Cr Magnetic Superlattices. Phys. Rev. Lett. \textbf{61}, 2472 (1988).

$^{3}$ Moodera, J. S., Kinder, L. R., Wong, T. M., Meservey, R. Large Magnetoresistance at Room Temperature in Ferromagnetic Thin Film Tunnel Junctions. Phys. Rev. Lett., \textbf{74}, 3273 (1995).

$^{4}$ Slonczewski, J. C. Current-driven excitation of magnetic multilayers, J. Mag. Mag. Mat. \textbf{159}, L1 (1996).

$^{5}$ Mangin, S., Ravelosona, D., Katine, J. A., Carey, M. J., Terris, B. D., Fullerton, E. E. Current-induced magnetization reversal in nanopillars with perpendicular anisotropy. Nat. Mater. \textbf{5}, 210 (2006).

$^{6}$ Kiselev, S. I., Sankey, J. C., Krivorotov, I. N., Emley, N. C., Schoelkopf, R. J., Buhrman, R. A., Ralph, D. C. Microwave oscillations of a nanomagnet driven by a spin-polarized current. Nature, \textbf{425}, 380 (2003). 

$^{7}$ Houssameddine, D., Ebels, U., Delaet, B., Rodmacq, B., Firastrau, I., Ponthenier, F., Brunet, M., Thirion, C., Michel, J.-P. , Prejbeanu-Buda, L., Cyrille, M. C., Dieny B. Spin-torque oscillator using a perpendicular polarizer and a planar free layer. Nat. Mater. \textbf{6}, 447 (2007). 

$^{8}$ Sbiaa, R. Magnetization switching by spin torque effect in off-aligned structure with perpendicular anisotropy. J. Phys. D: App. Phys. \textbf{46}, 395001 (2013).

$^{9}$ Rie Matsumoto, R., Arai, H., Yuasa, S., Imamura H. Spin-transfer-torque switching in a spin-valve nanopillar with a conically magnetized free layer. Appl. Phys. Express \textbf{8}, 063007 (2015).

$^{10}$ Nunn, Z. R., and Girt, E. U.S. Patent Application No. 15/919,071 (2018).

$^{11}$ Grunberg, P., Schreiber, R., Pang, Y., Brodsky, M. B., Sowers, H. Layered magnetic structures: Evidence for antiferromagnetic coupling of Fe layers across Cr interlayers. Phys. Rev. Lett. \textbf{57}, 2442 (1986).

$^{12}$ Heinrich, B., Celinski, Z., Cochran, J. F., Muir, W. B., Rudd, J., Zhong, Q. M., Arrott, A. S., Myrtle, K., Kirschner, J. Ferromagnetic and antiferromagnetic exchange coupling in bcc epitaxial ultrathin Fe (001)/Cu (001) Fe (001) trilayers. Phys. Rev. Lett. \textbf{64}, 673 (1990).

$^{13}$ Parkin, S. S. Systematic Variation of the Strength and Oscillation Period of Indirect Magnetic Exchange Coupling through the 3d, 4d, and 5d Transition Metals. Phys. Rev. Lett. \textbf{67}, 3598 (1991).

$^{14}$ Franke, P., Neuschutz, D. Thermodynamic Properties of Inorganic Materials · Binary Systems, Landolt-Börnstein - Group IV Physical Chemistry, Vol. 19B5 (2007).

$^{15}$ Predel B.  Phase Equilibria, Crystallographic and Thermodynamic Data of Binary Alloys, Landolt-Börnstein - Group IV Physical Chemistry, Vol. 5C (1993).

$^{16}$ Stiles, M. D. Interlayer exchange coupling. Ultrathin
Magnetic Structures III (pp. 99-142). Springer, Berlin, Heidelberg,
(2005).

$^{17}$ Eyrich, C., Zamani, A., Huttema, W., Arora, M., Harrison, D., Rashidi, F., Broun, D., Heinrich, B., Mryasov, O., Ahlberg, M. Karis, O., Jönsson, P. E., From, M., Zhu, X., Girt, E. Effects of substitution on the exchange stiffness and magnetization of Co films. Phys. Rev. B, \textbf{90}, 235408 (2014).

$^{18}$ Zhang, Z., Zhou, L., Wigen, P. E., Ounadjela, K. Angular
dependence of ferromagnetic resonance in exchangecoupled
Co/Ru/Co trilayer structures. Phys. Rev. B \textbf{50},
6094 (1994).

$^{19}$ Slonczewski, J. C. Conductance and exchange coupling
of two ferromagnets separated by a tunneling barrier. Phys.
Rev. B, \textbf{39}, 6995 (1989).

$^{20}$ de Vries, J. J., Kohlhepp, J., den Broeder, F. J. A., Coehoorn,
R., Jungblut, R., Reinders, A., de Jonge, W. J. M.
Exponential Dependence of the Interlayer Exchange Coupling
on the Spacer Thickness in MBE-grown Fe/SiFe/Fe Sandwiches.
Phys. Rev. Lett. \textbf{78}, 3023 (1997).

$^{21}$ Wang, Y., Levy, P. M., and Fry, J. L. Interlayer magnetic coupling in Fe/Cr multilayered structures. Phys. Rev. Lett. \textbf{65}, 2732 (1990).

$^{22}$ Shi, Z. P., Levy, P. M., and Fry, J. L.  Antiferromagnetic bias in the interlayer magnetic coupling. Europhys. Lett., \textbf{26}, 473 (1994).

$^{23}$ Shi, Z. P., Levy, P. M., and Fry, J. L. Interlayer magnetic coupling in metallic multilayer structures. Phys. Rev. B, \textbf{49}, 15159 (1994).

$^{24}$ Filipkowski, M. E., Krebs, J. J., Prinz, G. A., Gutierrez, C. J. Giant near-90 coupling in epitaxial CoFe/Mn/CoFe sandwich structures. Phys. Rev. Lett. \textbf{75}, 1847 (1995).

$^{25}$ Zoll, S., Van den Berg, H. A. M., Jay, J. P., Elmers, H. J., Meny, C., Panissod, P., Stoefflera, D.,  Diniaa, A., Ounadjela, K. Coupling mechanism in Co/Ru sandwiches with thin spacers. J. Magn. Magn. Mat. \textbf{156}, 231 (1996).

$^{26}$ Coey, J. E. D. Magnetism and Magnetic Materials. Cambridge University Press (2009).

$^{27}$ Barnas, J., Grunberg, P.  On the biquadratic interlayer coupling in layered magnetic structures. J. Magn. Magn. Mat. \textbf{121}, 326 (1993).

$^{28}$ Erickson, R. P., Hathaway, K. B., Cullen, J. R. Mechanism for non-Heisenberg-exchange coupling between ferromagnetic layers. Phys. Rev. B, \textbf{47}, 2626 (1993).

$^{29}$ Bobo, J. F., Kikuchi, H., Redon, O., Snoeck, E., Piecuch, M., White, R. L. Pinholes in antiferromagnetically coupled multilayers: Effects on hysteresis loops and relation to biquadratic exchange. Phys. Rev. B \textbf{60}, 4131 (1999).

$^{30}$ Slonczewski, J. C. Origin of biquadratic exchange in
magnetic multilayers. J. Appl. Phys. \textbf{73}, 5957 (1993).

$^{31}$ Slonczewski, J. C. Fluctuation mechanism for biquadratic exchange coupling in magnetic multilayers. Phys. Rev. Lett. \textbf{67}, 3172 (1991).

$^{32}$ Lee, Y. M., Hayakawa, J., Ikeda, S., Matsukura, F., Ohno, H.  Giant tunnel magnetoresistance and high annealing stability in CoFe/MgO/CoFeB magnetic tunnel junctions with synthetic pinned layer. Appl. Phys. Lett. \textbf{89}, 042506 (2006).

$^{33}$ Fullerton, E. E., Childress, J. R. Spintronics, Magnetoresistive Heads, and the Emergence of the Digital World. Proc. IEEE, \textbf{104}, 1787 (2016).

$^{34}$ Heinrich, B., Celinski, Z., Cochran, J. F., Arrott, A. S., Myrtle, K., Purcell, S. T., Bilinear and biquadratic exchange coupling in bcc Fe/Cu/Fe trilayers: Ferromagnetic-resonance and surface magneto-optical Kerr-effect studies. Phys. Rev. B \textbf{47}, 5077 (1993).

$^{35}$ Demokritov, S. O. Biquadratic interlayer coupling in layered magnetic systems. J. Phys. D: Appl. Phys. \textbf{31} 925 (1998).

\newpage

\newpage

\section{Extended Figure}
\renewcommand{\figurename}{EXTENDED FIG.}
\setcounter{figure}{0}

	\begin{figure}[tbph]
		\centering
		\includegraphics[width=0.49\textwidth]{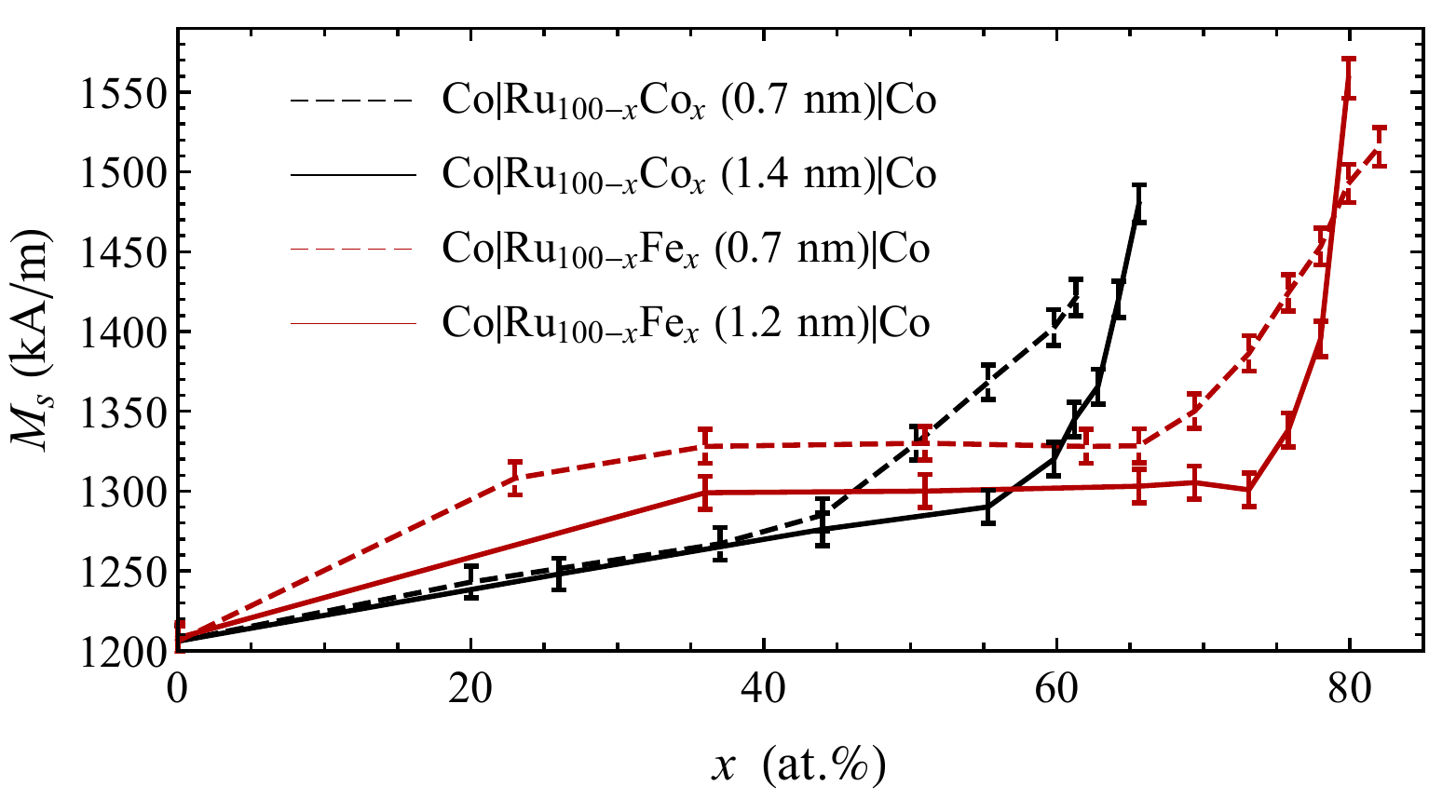}
		\caption{
			Total saturation magnetization, $M_{s}$, dependence on magnetic material concentration $x$ of four samples: Co(2 nm)$|$Ru$_{100-x}$Co$_{x}$(0.7 nm)$|$Co(2 nm), Co(2 nm)$|$Ru$_{100-x}$Co$_{x}$(1.4 nm)$|$Co(2 nm), Co(2 nm)$|$Ru$_{100-x}$Fe$_{x}$(0.7 nm)$|$Co(2 nm), and Co(2 nm)$|$Ru$_{100-x}$Fe$_{x}$(1.2 nm)$|$Co(2 nm).}
		\label{fig:MsAll}
	\end{figure}

\section{Methods}

Radio-frequency (rf) magnetron sputtering was used to deposit Ta(3.5)$|$Ru(3.5)$|$Co($t$)$|$Ru$_{100-x}$Co$_{x}$($d$)$|$Co($t$)$|$Ru(3.5) and Ta(3.5)$|$Ru(3.5)$|$Co($t$)$|$Ru$_{100-x}$Fe$_{x}$($d$)$|$Co($t$)$|$Ru(3.5) structures, and Ru$_{100-x}$Co$_{x}$(18) and Ru$_{100-x}$Fe$_{x}$(18) single films at room temperature on (100) Si substrates. In these structures, $x$ is the atomic concentration of Co or Fe in the Ru coupling layer, and the numbers in parentheses indicate the layer thicknesses in nm. The thickness, $d$, of RuCo and RuFe coupling layers was varied from 0.4 to 1.4 nm and the thickness, $t$, of Co was 2 and 8 nm. The Ta seed layer was used to induce the $\big \langle$0001$\big \rangle$ growth orientation of Ru, Co, Ru$_{100-x}$Co$_{x}$ and Ru$_{100-x}$Fe$_{x}$ layers. The role of the top Ru(3.5) is to protect Co layers from oxidation. 

Sputter deposition was performed at an argon pressure below 2 mTorr. The base pressure of the system was below 5$\times 10^{-8}$ Torr. Before deposition, (100) Si substrates were cleaned with the standard RCA SC-1 process.

A calibration of the growth rates was inferred from fitting X-ray reflectivity measurements of single layers of each material and alloy. X-ray diffraction measurements also showed that full structures have strong texture along the $\big \langle$0001$\big \rangle$ crystallographic orientations with a c-axis full-width-at-half-maximum distribution under 5$\degree$. The single Ru$_{100-x}$Co$_{x}$(18) and Ru$_{100-x}$Fe$_{x}$(18) films grown on Si substrates have hcp crystal structure and weak texture along $\big \langle$0001$\big \rangle$ crystal directions.

To determine the bilinear, $J_{1}$, and biquadratic, $J_{2}$, coupling constants, $M(H)$ hysteresis curves were collected by SQUID in up to a $\pm7$ T magnetic field. The coupling parameters and non-collinear coupling angle were determined by simulations using the micromagnetic model proposed by Eyrich et al.$^{17}$. The model assumes that each Co layer consists of $N$ sub-layers that interact via the direct exchange interaction, and the spins in each Co sub-layer rotate coherently. The coupling is established only between the two Co sub-layers adjacent to the Ru, RuCo, and RuFe coupling layers. The interlayer coupling is measured in $\big \langle 0001 \big \rangle$ textured Co$|$RuCo$|$Co and Co$|$RuFe$|$Co structures with Co layer thickness $t$ = 2 and 8 nm.  The uniaxial magnetocrystalline anisotropy field of Co layers is along the $\big \langle$0001$\big \rangle$ directions, perpendicular to the Co film plane. The demagnetizing dipolar field in Co films is much larger than the uniaxial magnetocrystalline anisotropy field forcing the magnetization to lie in the plane of the film. Due to the polycrystalline nature of the studied samples, the in-plane magnetocrystalline anisotropy is averaged. Then in the presence of an inplane external magnetic field, the anisotropy and demagnetizations energies can be ignored and the total magnetic energy, $E_{\text{Total}}$, can be written as 

\begin{equation}\label{total}
\begin{split}
&E_{\text{Total}}=J_{1}cos(\theta_{N}-\theta_{N+1})+J_{2}cos^{2}(\theta_{N}-\theta_{N+1})\\&+\dfrac{2A_{ex}}{a}\Bigg[ \sum_{i=1}^{N-1} cos(\theta_{i}-\theta_{i+1})+\sum_{i=N+1}^{2N-1} cos(\theta_{i}-\theta_{i+1})\Bigg]\\&+M_{\textbf{s}}Ha\sum_{i=1}^{2N} cos(\theta_{i}) ,
\end{split}
\end{equation}

where $A_{ex}$ the exchange stiffness, $a$ is the distance between atomic planes in Co, $N$ is the number of sublayers in the ferromagnetic layers, $H$ is the applied external field, and $M_{s}$ is the magnetic saturation of the Co layers. To fit the $M(H)$ curves, Eq.\ref{total} is minimized for individual Co sublayers. From this the total magnetization along the external magnetic field direction as a function of the field strength is calculated. The model has three fitting parameters $J_{1}$, $J_{2}$, and $A_{ex}$.  

The $A_{ex}$ was shown to vary with the thickness of Co layers$^{17}$ from 15.5 $\times$ 10$^{-12}$ J/m for $t$ $>$ 7 nm to 7.3 $\times$ 10$^{-12}$ J/m for $t$ = 2 nm if antiferromagnetic coupling between Co layers is described with only $J_{1}$. However, if both $J_{1}$ and $J_{2}$ are used to account for the antiferromagnetic coupling, the change of the $A_{ex}$ with $t$ is smaller, 15.5 $\times$ 10$^{-12}$ J/m for $t$ $>$ 7 nm to 10.5 $\times$ 10$^{-12}$ J/m for $t$ = 2 nm. This indicates correlation between the $A_{ex}$ and $J_{2}$ parameters in Co films with $t$ $<$ 7 nm. To account for this correlation, $M(H)$ curves are fit assuming the highest and lowest values, $A_{ex}$ = 0.9$\times$10$^{-11}$ and 1.7$\times$10$^{-11}$ J/m, of Co layers. The $J_2$ error bar accounts for this uncertainty in $A_{ex}$ values. For this reason the $J_2$ error bars are larger than those of $J_1$ in Fig.\ref{fig:J1J2phi}.

$\theta$ can be determined from minimizing the coupling energies or from the remnant magnetization, i.e., projection of magnetization along the applied magnetic field, H, direction at H = 0. For $\theta < 45\degree$, $\theta$ is more difficult to measure since the remnant magnetization is proportional to $\cos(\theta)$. Thus, the error bars on $\theta$ measurements increase as $\theta$ becomes smaller as is evident form Fig.\ref{fig:J1J2phi}c and Fig.\ref{fig:J1J2phi}f. 

With an increase of $J_1$ and $J_2$ the magnetic field, $H_s$, required to saturate the Co magnetic moments of Co(2)$|$Ru$_{100-x}$Co$_x$($d$)$|$Co(2) and Co(2)$|$Ru$_{100-x}$Fe$_x$($d$)$|$Co(2) also increase. $H_s$ of some measured structures approach 7 T, the magnetic field available in our SQUID magnetometer. In this case error bars of measured $J_1$ and $J_2$ also increase.
	
\end{document}